# e-G2C: A 0.14-to-8.31 µJ/Inference NN-based Processor with Continuous On-chip Adaptation for Anomaly Detection and ECG Conversion from EGM

Yang Zhao[1], Yongan Zhang[1], Yonggan Fu[1], Xu Ouyang[1], Cheng Wan[1], Shang Wu[1], Anton Banta[1], Mathews M. John[2], Allison Post[2], Mehdi Razavi[2], Joseph Cavallaro[1], Behnaam Aazhang[1], Yingyan Lin[1]*

[1]Rice University, Houston, Texas, USA; [2]Texas Heart Institute, Houston, Texas, USA; *Email: yingyan.lin@rice.edu

**Abstract**

This work presents the first silicon-validated dedicated EGM-to-ECG (G2C) processor, dubbed e-G2C, featuring continuous lightweight anomaly detection, event-driven coarse/precise conversion, and on-chip adaptation. e-G2C utilizes neural network (NN) based G2C conversion and integrates 1) an architecture supporting anomaly detection and coarse/precise conversion via time multiplexing to balance the effectiveness and power, 2) an algorithm-hardware co-designed vector-wise sparsity resulting in a 1.6-1.7× speedup, 3) hybrid dataflows for enhancing near 100% utilization for normal/depth-wise(DW)/point-wise(PW) convolutions (Convs), and 4) an on-chip detection threshold adaptation engine for continuous effectiveness. The achieved 0.14-8.31 µJ/inference energy efficiency outperforms prior arts under similar complexity, promising real-time detection/conversion and possibly life-critical interventions.

## Introduction

Globally, about 600K cardiac pacemakers are implanted annually. There has been a growing interest in remote pacemaker monitoring for reducing the frequency of costly hospital visits. Traditionally pacemakers utilize locally sensed EGM to provide therapy, but 12-channel ECG can provide more information for better diagnosis/therapy (Fig. 1) [1]. Presently there is no means to do G2C conversion in pacemakers and real-time on-pacemaker G2C conversion is highly desirable. While NN-based conversion offers a higher conversion quality and is more robust to real-world noise, artifacts, and diverse pathologies [1], it requires a prohibitive computational cost. To tackle this, e-G2C integrates 1) an architecture supporting a precise and more complex conversion algorithm in instant response to only the detected anomaly (i.e., arrhythmia) and activates a coarse conversion otherwise to balance the effectiveness and power consumption via time multiplexing (**Fig. 1**); 2) algorithm-hardware co-design for enabling sparse NN processing; and 3) hybrid dataflows for enhancing near 100% utilization for all Convs. Furthermore, e-G2C is equipped with 4) a continuous on-chip learning engine to adapt the detection threshold over time for each patient.

## Proposed e-G2C Processor

**1. Overall Architecture:** e-G2C includes two main engines: NN and adaptation engines (**Fig. 2**). In particular, the NN engine consists of weight, index, input/output activation (Act) memories, and 32 multiplication-and-accumulation (MAC) lanes. To orchestrate the processing of the three different models including anomaly detector and coarse/precise convertors, the on-chip controller reads 32-bit instructions from the instruction SRAM to control the processor. First, a weight global buffer (GB) stores the compressed sparse weights, and an index SRAM stores the weight indices, while each MAC lane performs row-wise intra-channel reuse (RIR) that exists in all Convs. Second, e-G2C adopts two Act GBs that iterate layer-wise as input/output Act GB to eliminate off-chip Act accesses. Third, to pipeline the data preparation and computation, an input and output Act buffer are inserted. Finally, the weight buffer supports both 4-bit power-of-2 and 8-bit int weight formats to provide coarse/precise processing. The adaptation engine consists of a comparator (Cmp.), an argmin calculator (Argmin), and registers for the adaptation.

**2. Algorithm-hardware Co-designed Vector-wise Sparse NN Processing: Fig. 3** shows e-G2C's algorithm-hardware co-designed vector-wise sparsity for both normal and PW Conv and a supporting input Act buffer design. On the algorithm level, one sparse vector corresponds to one row of weights in normal Conv and to three weights in PW-Conv. Such vector-wise sparsity offers the benefits of skipping the computations and thus reduces latency and energy. On the hardware level, the input Act buffer temporally stores input Acts, and uses indices to select proper Acts for each MAC lane.

**3. Intra-channel Reuse for DW-Conv:** Since Acts in DW-Conv cannot be reused by different weight kernels, e-G2C leverages the reuse opportunities of Acts for one weight kernel, i.e., column-wise intra-channel reuse (CIR) and deeper RIR (D-RIR) (**Fig. 4**), for boosted utilization. In CIR, one input Act row is reused by three weight rows of one column for its corresponding three output Act rows of one column. e-G2C maps the weight rows and the corresponding input Act row into three MAC lanes to increase the utilization by 3×. In D-RIR, e-G2C divides one input Act row into two sub rows, corresponding to two sub rows of length-4 output Act in one output row, and further spatially maps these computations to two MAC lanes for 2× utilization. These hybrid dataflows for different Convs result in near 100% utilization for all Convs. **Fig. 5** lists the models' structures and accuracies on a clinically collected dataset and the speedup with the proposed techniques.

**4. The Anomaly Detection:** For anomaly detection, the detection threshold needs to adapt to the heart condition over time and across patients, requiring continuous threshold adaptation for real-world effectiveness. As an optimal anomaly threshold aligns with the detection output within a sensitive range, our e-G2C chip integrates a threshold adaptation flow consisting of three steps (**Fig. 6**): (1) calculating the histogram of the detector's outputs for T days (e.g., T = 3) by using Cmp., (2) selecting the interval with the least occurrence based on the histogram through Argmin, and (3) setting the mean of the above interval as the threshold. Such threshold adaptation leads to a 3.45%-4.00% higher accuracy against a static threshold.

## Measurement Results

e-G2C is fabricated in 28nm CMOS technology (**Fig. 7**), and **Fig. 8** summarizes measurement results together with related prior works. The achieved energy efficiency for detection (0.14 µJ/detection) and conversion (4.13 µJ/conversion and 8.31 µJ/conversion for coarse/precise conversion, respectively) outperforms prior arts with similar model complexity. The achieved latency for one coarse/precise conversion with detection is 1.32%/2.66% of one heartbeat period (assuming a heart rate of 80 bpm), promising real-time detection/conversion and possibly life-critical interventions.

**Acknowledgement.** This work was supported by National Institutes of Health (R01HL144683), National Science Foundation (1937592), and Silicon Creations.

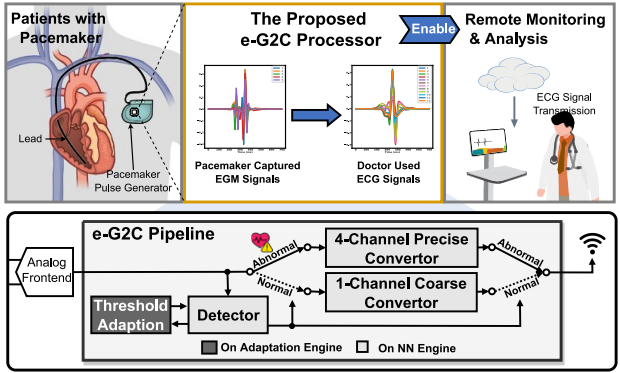

Fig. 1 Visualizing the proposed e-G2C pipeline.

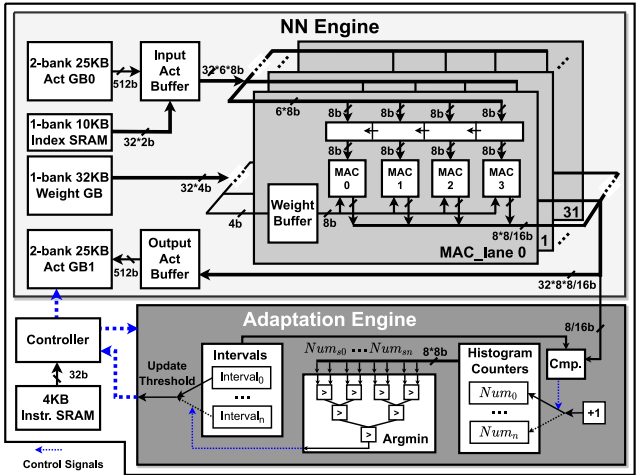

Fig. 2 e-G2C overall architecture.

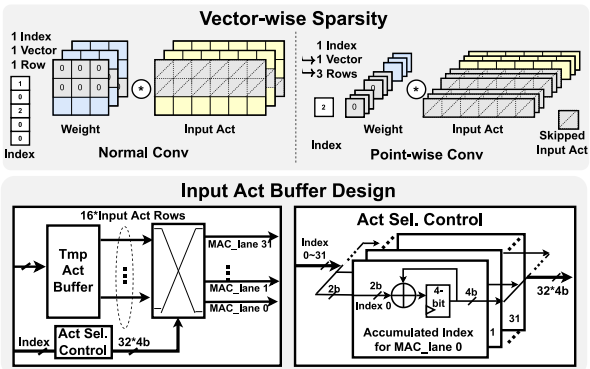

Fig. 3 e-G2C processor's adopted vector-wise sparse algorithm and the corresponding activation buffer design.

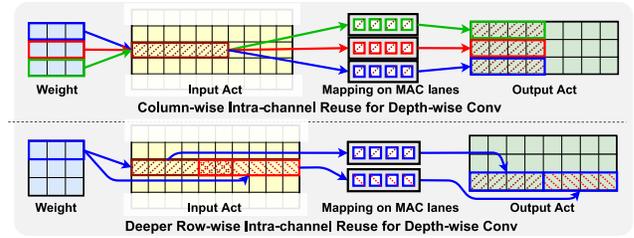

Fig. 4 The proposed intra-channel reuse dataflow for DW Conv.

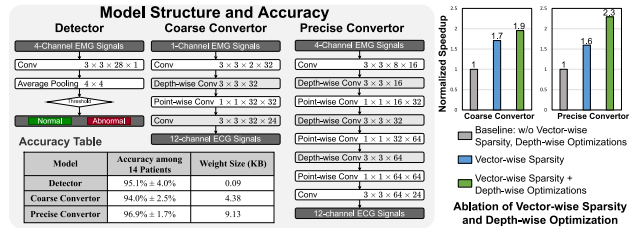

Fig. 5 The three models' structures and accuracy.

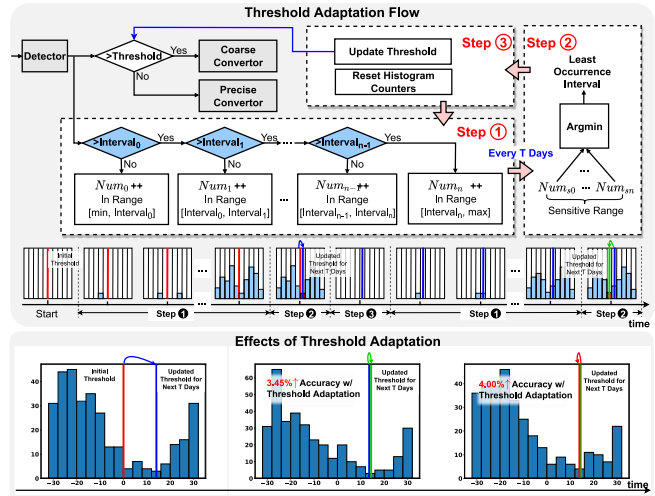

Fig. 6 The on-chip anomaly threshold adaptation flow and an illustration of an example achieved a 3.45%-4.00% higher accuracy.

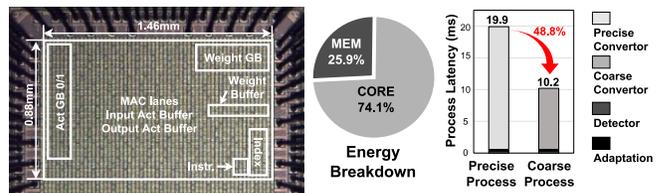

Fig. 7 Chip photograph, energy breakdown, and measured latency.

| | ISSCC'21 [2] | TBIOCAS'20 [3] | JSSC'20 [4] | JSSC'18 [5] | TBIOCAS'19 [6] | JETC'21 [1] | This Work |
|---|---|---|---|---|---|---|---|
| Process | 65 nm | 180 nm | 40 nm | 130 nm | N/A | Xilinx AC706 | 28 nm |
| Core Area | 1.75 mm² | 0.93 mm² | 4.54 mm² | 3.3 mm² | N/A | N/A | 1.3 mm² |
| Supply Voltage | 0.75 V | 1.8 V | 0.58 V | 1.2 V | N/A | N/A | 0.30 V (CORE) 0.59 V (MEM) |
| Memory | 73 KB | 8 KB | 35.8 KB | 96 KB | N/A | N/A | 104 KB |
| Frequency | 1MHz/2.5MHz | 25MHz | 65KHz/130KHz | 10MHz | N/A | 200MHz | 2MHz |
| Bit Precision | 16-bit | 16-bit | 12/24-bit | 16-bit | N/A | 8~16~32-bit | 4-/8-bit (Weight) 8-bit (Activation) 8-/16-bit (Output) |
| Support MLP | Yes | Yes | No | No | Yes | Yes | Yes |
| Support Conv Type — Normal | Yes | No | No | No | Yes | Yes | Yes |
| Support Conv Type — DW | No | No | No | No | No | No | Yes |
| Support Conv Type — PW | No | No | No | No | Yes | Yes | Yes |
| Task | ECG Classification | ECG Classification | Seizure Detection | Seizure Detection | ECG Classification | EGM-to-ECG Conversion | EGM Classification EGM-to-ECG Conversion |
| Computational Complexity [1] | N/A | 3.6 K | N/A | 123.9 K | 164 K | 28.87~31.38 M | Detector: 4 K Coarse Convertor: 2.69 M Precise Convertor: 5.79 M |
| On-chip Adaptation | Yes | No | Yes | No | No | No | Yes |
| Power | 46.8µW @ 1MHz 86.7µW @ 2.5MHz | 13.34 µW | 1.9 mW | 670 µW | 134 mW [2] | 10-15 W | 430.0 µW |
| Energy Efficiency | 2.25 µJ/Classification | 3.21 µJ/Classification | 170.9 µJ/Detection | 168.6 µJ/Detection | N/A | 1440-4370 µJ/Conversion [3] | 0.14 µJ/Detection 4.13 µJ/Conversion (Coarse) 8.31 µJ/Conversion (Precise) |
| Latency | 31 ms/Classification | 0.25 ms/Classification | 710 ms/Detection | <1000 ms/Detection | N/A | 1.44-4.37 ms/Conversion [3] | 0.32 ms/Detection 9.62 ms/Conversion (Coarse) 13.32 ms/Conversion (Precise) |

1 Use number of operations of one algorithm to quantify computational complexity. Each MAC is one operation
2 Estimated.
3 Estimated according to the frame rate.

Fig. 8 Chip measurements and comparison with recent prior works.